# Evolution of precipitate morphology during heat treatment and its implications for the superconductivity in K$_x$Fe$_{1.6+y}$Se$_2$ single crystals

Y. Liu[*], Q. Xing, K. W. Dennis, R. W. McCallum, and T. A. Lograsso

*Division of Materials Sciences and Engineering, Ames Laboratory,*

*US DOE, Ames, Iowa 50011, USA*

We study the relationship between precipitate morphology and superconductivity in K$_x$Fe$_{1.6+y}$Se$_2$ single crystals grown by self-flux method. Scanning electron microscopy (SEM) measurements revealed that the superconducting phase forms a network in the samples quenched above iron vacancy order–disorder transition temperature $T_s$, whereas it aggregates into micrometer-sized rectangular bars and aligns as disconnected chains in the furnace-cooled samples. Accompanying this change in morphology the superconducting shielding fraction is strongly reduced. By post-annealing above $T_s$ followed by quenching in room temperature water, the network recovers with a superconducting shielding fraction approaching 80% for the furnace-cooled samples. A reversible change from network to bar chains was realized by a secondary heat treatment in annealed samples showing a large shielding fraction, that is, heating above $T_s$ followed by slow cooling across $T_s$. The large shielding fraction observed in K$_x$Fe$_{1.6+y}$Se$_2$ single crystals actually results from an uniform and contiguous distribution of superconducting phase. Through the measurements of temperature dependent x-ray diffraction, it is found that the superconducting phase precipitates while the iron vacancy ordered phase forms together by cooling across $T_s$ in K$_x$Fe$_{1.6+y}$Se$_2$ single crystals. It is a solid solution above $T_s$, where iron atoms randomly occupy both Fe1 and Fe2 sites in the iron vacancy disordering status; and phase separation is driven by the iron vacancy order–disorder transition upon cooling. However, neither additional iron in the starting mixtures nor as-quenching at high temperatures can extend the miscibility gap to the KFe$_2$Se$_2$ side.

**PACS number(s):** 74.70.Xa, 74.81.Bd, 74.25.F-

[*] Corresponding author: yliu@ameslab.gov

**I. INTRODUCTION**

Since the pioneering work on superconductivity in iron oxypnictides reported by Hosono's group [1-3], research on iron-based superconductors has greatly increased. The latest discovered $A_xFe_{2-y}Se_2$ (A=K, Rb, Cs, and Tl) superconductors spur a new gold rush in the field [4]. In contrast to formerly discovered "1111", "122", and "11" compounds etc., the coexistence of two spatially-separated phases is observed in $K_xFe_2Se_2$ superconductors by high-resolution transmission electron microscopy (HRTEM) [5-6], as well as scanning nanofocused x-ray diffraction [7-8]. The unusual features observed in optical spectroscopy [9] and angle-resolved photoemission spectroscopy (ARPES) measurements [10] were explained by the phase separation scenario. The muon spin rotation (μSR) [11], Mössbauer [12], and nuclear magnetic resonance (NMR) [13] experiments have further revealed nearly 90% of the sample volumes exhibit large-moment antiferromagnetic (AFM) order, while 10% of the sample volumes remain paramagnetic (PM) and attributed to a metallic/superconducting phase in $A_xFe_{2-y}Se_2$ single crystals. Detailed structure analysis revealed that the main phase has a $\sqrt{5}\times\sqrt{5}\times1$ superstructure, where Fe1 sites are fully occupied and the Fe2 sites are almost empty [14-16]. The iron vacancy order–disorder transition takes places above 500 K, followed by an AFM order with large saturated moment (~3.3 $\mu_B$ per Fe at 10 K) [15-16]. It is very important to identify the superconducting phase in $A_xFe_{2-y}Se_2$ single crystals. The scanning tunneling microscope (STM) measurements demonstrated that the superconducting phase has a complete FeSe layer or with Se vacancy defects, close to the ideal 122 structure [17-18]. NMR experiments also suggested the superconducting phase in $Rb_{0.74}Fe_{1.6}Se_2$ single crystal to be $Rb_{0.3(1)}Fe_2Se_2$ [13]. In $A_xFe_{2-y}Se_2$ single crystals showing two transitions at 30 and 44 K, a special region with a grain size of 60 μm has been recognized with a ratio of $K_{0.9}Fe_{2.17}Se_2$ [19]. Therefore, $A_xFe_{2-y}Se_2$ compounds still follow the same mechanism as other iron arsenides or chalcogenides, where FeAs (FeSe) layers and $FeAs_4$ ($FeSe_4$) tetrahedra are responsible for superconductivity. More recently, it was discovered that superconductivity at $T_c$ = 46 K can be obtained by intercalating metals, Li, Na, Ba, Sr, Ca, Yb, and Eu between FeSe layers by synthesizing the compounds in liquid ammonia [20-21]. Crystal structure analysis suggests $Fe_2Se_2$ layers are separated by lithium ions, lithium amide, and ammonia

[21].

To date, $A_x\text{Fe}_{2-y}\text{Se}_2$ single crystals have been grown by different techniques, such as the self-flux method, Bridgman technique, and optical floating zone technique [22-32]. We notice that the post-annealing and quenching technique can improve the superconductivity of non-superconducting as-grown crystals [33]. Annealed crystals show a nearly full shielding fraction and high critical current density $J_c$ [34-35]. Liu *et al*. reported that the superconducting $K_x\text{Fe}_{2-y}\text{Se}_2$ single crystals had been grown using the optical floating zone technique, where the lamp power was switched off immediately towards the end of the crystal growth, resulting in a fast cooling rate [32]. However, it is not well understood how the superconductivity can be controlled during the growth process and why the quenching technique/fast cooling yields samples with a large superconducting shielding fraction. Triggered by the elucidation of mechanism of phase separation and the formation of the superconducting phase, a series of single crystals of nominal compositions $K_{0.8}\text{Fe}_{2+z}\text{Se}_2$ ($z$=0, 0.2, and 0.6) have been grown by a self-flux method by as-quenching, furnace-cooling, and post-annealing treatments. As-quenched samples show large shielding fraction ranging 20%-50%, whereas the shielding fraction drops below 1% in furnace-cooled crystals. For the latter, however, the shielding fraction can be greatly enhanced by post-annealing followed by quenching. The scanning electron microscopy (SEM) was applied to study the phase separation in $K_x\text{Fe}_{1.6+y}\text{Se}_2$ single crystals. It is found that the superconducting phase forms a network in as-quenched and post-annealed samples, whereas it aggregates into disconnected grain chains, referred to as bar chains in order to distinguish the network pattern. The results demonstrate that the large shielding fraction results from a uniform and contiguous distribution of the superconducting phase. Through the measurements of temperature dependent x-ray diffraction (XRD) and differential scanning calorimetry (DSC), we present a picture to describe the formation of exsolution texture in $K_x\text{Fe}_{1.6+y}\text{Se}_2$ single crystals.

**II. EXPERIMENTAL DETAILS**

A FeSe precursor, consisting of iron powder (99.998 wt.% purity) and selenium shot (99.999 wt.% purity), was synthesized at 973 K for 12 hours. Potassium was weighed at an atomic ratio of K:Fe:Se=0.8:2:2, the mixture was subsequently heated to 1023 K for 6 hours.

The chemicals were loaded into an alumina crucible and sealed in a quartz tube under a partial argon atmosphere. A $K_{0.8}Fe_2Se_2$ polycrystalline sample was thoroughly ground into powder by adding additional iron at ratios of 0.8:2+$z$:2 ($z$=0, 0.2, and 0.6). The powder was sealed in a small quartz ampoule with a diameter of 0.8 cm. This small ampoule was then sealed in a bigger ampoule. The crystal growth was performed in a vertical tube furnace. The preparation and heat treatment conditions are summarized in Table I. Furnace-cooled crystals were obtained by natural cooling down to room temperature in the furnace. The ampoule was heated to 1303 K and held for 2 hours. The furnace was cooled down to 1053 K at a cooling rate of 6 K/h, and subsequently cooled down to 373 K during a period of 12 hours by switching off the electric power. Annealing treatments were carried out at 673, 773 and 873 K and held for 2, 4, and 6 hours at each temperature, followed by quenching in room temperature water. In addition, the furnace–cooled samples with starting compositions $K_{0.8}Fe_{2+z}Se_2$ ($z$=0.2 and 0.6) were just heated up to 573 K for 10 minutes and subsequently quenched in water. For as-quenched samples the ampoules were heated to 1403 K and cooled down to 1173 and 1073 K. Before quenching in room temperature water the ampoules were held at 1173 and 1073 K for 12 hours and 20 hours, respectively.

XRD measurements were performed on a PANalytical MPD diffractometer equipped with a high temperature stage using Co $K\alpha$ radiation. Samples were protected in an atmosphere of flowing helium gas during measurements. The $K\alpha2$ radiation was removed with X'pert Highscore software. The microstructure of $K_xFe_{1.6+y}Se_2$ single crystals was analyzed with JEOL JSM-5910LV and JXA-8200 microscopes using backscattered electron and secondary electrons. The actual compositions of crystals were determined by wave dispersive x-ray spectroscopy (WDX). DSC measurements were completed using a Netzsch DSC system by heating and cooling cycles at a rate of 10 K/min. $K_xFe_{1.6+y}Se_2$ single crystals with a mass of 40 mg were sealed in a tantalum capsule by laser welding. In-plane resistivity measurements were performed on a Physical Property Measurement System (PPMS, Quantum Design). Magnetic susceptibility was measured by using SQUID magnetometer (MPMS, Quantum Design) and PPMS Vibrating Sample Magnetometer (PPMS VSM, Quantum Design).

## III. RESULTS AND DISCUSSION

Figure 1 shows the optical picture of the ingot grown from a nominal composition of $K_{0.8}Fe_{2.6}Se_2$. Along the growth direction (long edge), the ingot can be readily cleaved with a blade from the upper to middle part. Amorphous matter sank in the cone-shaped bottom. Right top and bottom photos in Fig. 1 show the cleaved crystals with a shiny surface after annealing at 873 K for 4 hours. The actual composition was determined as $K_{0.81}Fe_{1.70}Se_2$ for the crystals cleaved from the upper part of the ingot, assuming no selenium deficiency in the crystals. The actual composition of the crystals cleaved from the middle part is almost the same as that in the upper part. However, the actual compositions show a large fluctuation from spot to spot in different areas. The iron content fluctuates such that K:Fe:Se=(0.51-0.78):(12.04-56.34):2 in the amorphous cone. The actual composition of the obtained crystals is $K_{0.85}Fe_{1.67}Se_2$ for the starting composition $K_{0.8}Fe_2Se_2$. It is a weak tendency that for more iron in the starting materials, the more iron in the obtained crystals.

Figure 2(a) shows the XRD spectra of the furnace-cooled and post-annealed crystals measured at room temperature. The peak separation of (00$l$) reflections evidences the coexistence of two spatially separated phases in the crystals. The minority phase is indicated by the asterisks in Fig. 2(a). The temperature dependent behavior of (008) reflection was recorded in heating [Fig. 2(b)] and cooling [Fig. 2(c)] cycles. As can be seen, the majority phase, that is, AFM ordered superstructure, and the shoulder, that is, the superconducting phase, are getting close with an increase of temperature and merge into one peak above 537 K. Upon cooling the superconducting phase appears again. The shoulder steadily shifts toward the low angle region, while the reflection of the majority phase stands still. The evolution of the $c$-axis lattice parameters is shown in Fig. 2(d). It is noticed that $c$-axis lattice parameter shows an increase below 541 K upon warming although the separation of two reflections can be identified below 537 K. A thermal hysteresis behavior ~2 K is observed during the heating and cooling cycles. We define $T_s$=541 K as the temperature where phase separation takes place, which is close to the iron vacancy order-disorder transition temperature revealed by neutron scattering measurements [15-16]. The result clearly evidences that phase separation takes place at $T_s$.

Figure 3(a) shows the temperature dependence of the magnetic susceptibility of the crystal obtained by as-quenching from 1173 K with a starting composition of $K_{0.8}Fe_{2.6}Se_2$. The shielding fraction fluctuates between 20% and 50% from the pieces cleaved in different areas. However, the shielding fraction drops below 1% in furnace-cooled crystals from the same starting composition, as shown in Fig. 3 (b). Interestingly, the shielding fraction can be greatly enhanced by post-annealing at 673 and 873 K and subsequently quenching in room temperature water. It is found that $T_c$ stabilizes at 32 K while shielding fraction approaches 80% in the crystals from different batches. With increasing iron content in the starting materials $K_{0.8}Fe_{2+z}Se_2$ ($z$=0, 0.2, and 0.6), the shielding fraction increases from 40% to 80% in the annealed $K_xFe_{1.6+y}Se_2$ single crystals. Our result is consistent with the previous reports, where the superconductivity in $A_xFe_{2-y}Se_2$ single crystals can be improved by increasing the iron content in the starting materials [29,31,36]. Figure 3(c) shows the normal-state magnetic susceptibility $\chi(T)$ of $K_xFe_{1.6+y}Se_2$ single crystals obtained by as-quenching, furnace-cooling and post-annealing techniques, respectively. With decreasing temperature, $\chi(T)$ gradually decrease and show the similar temperature-dependent behavior for all the samples.

In order to gain a better understanding of the dependence of $\chi(T)$ on thermal history, it is necessary to investigate the microstructural changes in the crystals grown by as-quenching, furnace-cooling and processed by post-annealing. Figure 4 shows clear evidence of coexistence of two phases in $K_xFe_{1.6+y}Se_2$ single crystals. SEM images were taken in the *ab* plane. For the crystals as-quenched at 1073 K, a modulated texture with a bright contrast is visible as the network, as shown in Fig. 4(a). This network is weaved by stripes perpendicular to each other. Each individual stripe consists of regularly ordered small rectangular bars. A similar stripe pattern has been reported previously [37-40]. In order to determine the orientation of the stripe, the orientation of the same piece was determined using back Laue x-ray diffraction. The measured back Laue pattern is given in the inset of Fig. 4(a), and the simulated one in Fig. 4(b). It is clear that strips orient along the [110] and [1$\bar{1}$0] directions, consistent with previous reports [37-38,40]. Here, the strips rotate 45° from the crystallographic *a* axis of AFM ordered $\sqrt{5}\times\sqrt{5}\times1$ superstructure. Based on this result we present a schematic drawing of the network pattern in the *ab* plane, as shown in Fig. 4(c). The superconducting phase consists of complete FeSe structure without iron vacancy, where the iron vacancy sites (Fe2 sites) are filled with red spots. It should be emphasized that the lattice distortions should exist at the phase interface, which are caused by different lattice parameters and shrinkage with varying temperature [39]. The orientation of strips is suggested to be aligned along the soft elastic strain direction. In fact, the strips are parallel to the edge of the square defined by the four closest iron vacancy sites. As these sites are occupied by excess iron atoms, lattice distortions can be relieved by stretching along this short way, instead of the diagonal direction of the iron vacancy square. A cartoon on the iron vacancy disordering status (DOS) is shown in Fig. 4(d). Upon warming, the iron vacancy ordered structure (OS) enters into disordering status above $T_s$, where iron atoms randomly occupy both Fe1 and Fe2 sites. Meanwhile, a superconducting phase with a 122 structure decomposes controlled by a fast diffusion of iron atoms in FeSe layers.

Figures 4(e) and 4(f) present the SEM images of the crystals as-quenched at 1173 and 1073 K, respectively. The average wavelength of the modulations as estimated from the micrographs is ~2 μm for the samples as-quenched at 1173 K. The wavelength of samples quenched at 1073 K slightly increases and the network becomes less dense, which show a smaller shielding fraction than samples quenched at 1173 K.

Intriguingly, network aggregates into disconnected strings of bars in the furnace-cooled samples grown from the nominal compositions of $K_{0.8}Fe_{2+z}Se_2$, as shown in Figs. 4(g) $z=0$, 4(h) $z=0.2$, and 4(i) and 4(j) $z=0.6$. We also find that the size of the bars of the furnace-cooled samples is nearly two or three times larger than those of the samples quenched at 1173 K by comparing Fig. 4(j) with Fig. 4(e). A WDX line scan along the bar chain reveals that the potassium and iron contents oscillate with the period well matching small bars passed by. The composition analysis suggests that the bars contain more iron and less potassium; whereas the dark matrix shows less iron and more potassium, consistent with the superconducting phase $Rb_{0.3(1)}Fe_2Se_2$ revealed by NMR measurements [13]. The shielding fraction is smaller than 1% for all the furnace-cooled crystals with different the starting materials $K_{0.8}Fe_{2+z}Se_2$ ($z=0, 0.2,$ and 0.6).

By post-annealing at 673 and 873 K for 2, 4, and 6 hours and subsequently quenching in room temperature water, the network recovers, as shown in Figs. 4(k)-(n). The shielding fraction is greatly enhanced (see Fig. 3). It is found that the size of the network cell becomes smaller with an increased annealing time. The average size of the network cell is ~ 2 μm between two knots for the samples annealed for 2 hours as well as as-quenched samples whereas it becomes finer at around 1 μm for the sample annealed for 6 hours [see Figs. 4(m) and 4(n)]. A secondary heat treatment was performed on these annealed samples showing sharp transition and large shielding fraction. By heating up to 873 K followed by furnace cooling down to room temperature, the network aggregates into bar chains, as shown in Figs. 4(o) and 4(p). As for the $c$ dimension of the superconducting phase, it was found that the height of the bar chains\network along the $c$-axis is around 5 nanometers, according to the HRTEM [5] and scattering-type scanning near-field optical microscopy (s-SNOM) measurements [40]. The superconducting phase is a thin plate shape, which is embedded in iron vacancy\AFM ordered phase and forms a lamellar structure along the $c$-axis direction [5,40].

Previously, an electronic and magnetic phase diagram of $K_xFe_{2-y}Se_2$ was reported as a function of iron valence by calculating the iron valence with the formula (4-$x$)/(2-$y$) [36]. Occurrence of superconductivity was suggested to be a sudden change of the Fermi surface at the boundary between the superconducting phase and the AFM semiconducting phase [36]. It was also suggested that the increased occupancy of the high-symmetry Fe2 site in the quenched sample is the key structural parameter that governs the bulk superconductivity [41]. Our results clearly evidence that the large shielding fraction in the annealed samples can be attributed to the formation of the network, where AFM ordered domains are shielded by the PM metallic phase. Such a network can be treated as two-dimensional arrays of Josephson junctions, which supports a Josephson-coupling plasmon observed by optical spectroscopy measurement [9]. However, it should be pointed out that a coupling between the superconducting phase and iron vacancy\AFM ordered phase [42] as well as lattice distortions along the $c$ axis [40] might have an influence on superconductivity as the precipitate morphology changes from bar chains to network. For the furnace-cooled samples, the superconducting phase clusters in disconnected strings of large grains, leading to low shielding fraction.

To clarify the mechanism of intergrowth in $A_xFe_{2-y}Se_2$ superconductors, DSC measurements were completed to probe the characteristic temperatures during the heat treatment and crystal growth process. Figure 5(a) shows the DSC curve of $K_xFe_{1.6+y}Se_2$ single crystals grown from the nominal composition of $K_{0.8}Fe_{2.6}Se_2$. The sample was sealed in a tantalum capsule by laser welding. During the heating process, an endothermic reaction occurred at 544 K, while an exothermic reaction occurred at 534 K upon cooling. The temperature 544 K matches the iron vacancy order–disorder transition temperature $T_s$=541 K measured by high-temperature x-ray measurements. A similar DSC peak has been observed in $Rb_xFe_{2-y}Se_2$ samples [39] and $Cs_xFe_{2-y}Se_2$ [43]. There are no endothermic or exothermic reactions observed at higher temperatures by examining several of the heating and cooling cycles. With a further increase in temperature to the maximum temperature applied during crystal growth, the crystal starts to melt and decompose at 1175 K, as shown in Fig. 5(b). This temperature is comparable to the molten zone temperature ~ 1162 K for crystal growth by utilizing the optical floating zone technique [32].

Now, we present a picture on the mechanism of exsolution intergrowth in $K_xFe_{1.6+y}Se_2$ superconductors. A schematic drawing of crystal growth is shown in Figs. 5(c)-(e). With an increase in temperature, the polycrystalline samples begin to melt and decompose above 1175 K. A higher soaking temperature of 1403 K was applied to dissolve the additional iron, $z$, at a risk of loss of potassium as we grew as-quenched samples. However, the starting mixtures could not be completely melted even at $T$=1403 K. Upon cooling, the iron-rich multiple components sink to the bottom of the ingot during crystal growth, as shown in Fig. 5(c). The crystallization ends at 1175 K, as revealed by DSC measurements. Quenching treatment above $T_s$ gives rises to the network pattern, whereas the superconducting phase aggregates into bar chains by slowly cooling across $T_s$.

It should be pointed out that the regularly arranged network results from exsolution intergrowth in a solid solution subject to soft directions of elastic strains. This modulated texture can be formed by diffusion-controlled nucleation and growth in a miscibility gap in the solid solution [44]. In the case of $K_xFe_{1.6+y}Se_2$ single crystals, a charge balance is realized in an ideal iron-vacancy-ordered phase in a formula of $K_{0.8}Fe_{1.6}Se_2$. A donation of 0.5 electrons per iron gives rise to instability of $KFe_2Se_2$ because of unusual $Fe^{1.5+}$ valence. To host the excess iron, another way is to reduce the amount of potassium atoms. Above $T_s$, iron atoms randomly occupy both Fe1 and Fe2 sites in the iron vacancy disordering status.

Additional iron in the starting materials is supposed to increase the amount of iron atoms which occupy the iron vacant sites in the parent compound, $K_{0.8}Fe_{1.6}Se_2$. Even though $K_xFe_{1.6+y}Se_2$ single crystals can host a small amount of iron impurity atoms during crystal growth at high temperature, the solubility of the iron impurity atoms ~$y$ is limited by the valence rule. Here, iron vacancy disordering status corresponds to the solid solution above $T_s$ [see Fig. 4(d)]. It should be emphasized that the iron vacancy order–disorder transition occurs in the $K_{0.8}Fe_{1.6}Se_2$ phase [45]. With decreasing temperature across $T_s$, the iron vacancy ordered structure is established. The iron impurity atoms in $K_{0.8}Fe_{1.6+y}Se_2$ solid solution are expelled from iron vacancy ordered area. The superconducting phase with the 122 structure precipitates and grows into rectangular bars, as shown in Fig. 4(c). The size and orientation of the rectangular bars are determined by the lattice strains and interface energy. The precipitate nucleation process, that is, phase separation, is driven by the iron vacancy order–disorder transition.

Based on above phase separation scenario, the furnace-cooled samples were simply heated to 543 K for 10 minutes, followed by quenching in water. Magnetic susceptibility $\chi(T)$ displays the same sharp transition and large shielding fraction as we observed in the samples quenched at 673 K and 873 K [see Fig. 3(b)]. Let's go back to quenched samples. Quenching above $T_s$ produces the network pattern, which just means that the quenching treatment does not completely freeze the iron vacancy disordering status. Instead of the expected homogeneous solid solution, the iron vacancy ordered structure has been broken up and divided into small domains, assuming a large diffusion coefficient in the iron vacant FeSe planes. By slowly cooling across $T_s$ the superconducting phase grows into large grains as we observed in the furnace-cooled samples. The quenching treatment itself should not create the superconducting phase by freezing the iron vacancy disordering status.

Figure 6(a) shows the temperature dependence of in-plane resistivity $\rho_{ab}$ of the crystals obtained by as-quenching, furnace-cooling, and post-annealing from the starting mixture $K_{0.8}Fe_{2+z}Se_2$ ($z=0.6$). For the furnace-cooled sample $\rho_{ab}$ exhibits a large hump with the maximum at $T_p$~120 K. Below $T_p$, $\rho_{ab}$ shows a metallic behavior, and a sample enters into the superconducting state at $T_c$~32 K. For the samples with a large shielding fraction such as as-quenched and post-annealed samples $T_p$ shifts towards high temperature and normal state resistivity decreases. Our results suggest that the resistivity hump relies on the ways of connection between the two electronic transport channels, that is, semiconducting (iron-vacancy ordering) and metallic\superconducting phases, consistent with the microstructures revealed by SEM measurements. Figures 6(b) and 6(c) show the resistivity broadening curves of post-annealed $K_xFe_{1.6+y}Se_2$ single crystals with the magnetic field applied parallel to the *c* axis ($H/\!/c$) and with field within the *ab* plane ($H/\!/ab$) of the sample. The anisotropic broadening behavior is similar to that reported earlier [32].

Figure 7 shows the magnetic hysteresis loops (MHLs) of post-annealed $K_xFe_{1.6+y}Se_2$ single crystals for (a) *H//c* and (b) *H//ab* at different temperatures. As can be seen, in the case of *H//c*, MHLs display a symmetric shape, while MHLs tilt for *H//ab*. As we discussed above, $A_xFe_{2-y}Se_2$ superconductors are not real bulk superconductors, although the shielding fraction appears large. Some groups reported that MHLs show a dip near the zero field instead of a central peak at temperatures below 10 K [30,46]. This feature is related to the phase separation nature of the samples [46]. Here, it should be pointed out that the abnormal dip is a sample-dependent phenomenon. The occurrence of this feature depends on the uniform and contiguous distribution of the superconducting phase.

Assuming the extended Bean critical state model [47], the field dependence of critical current density, $J_c$, has been estimated by the formula $J_c=20\Delta M/[w(1-w/3l)$ for *H//c*, where *l* is the length and *w* is the width of the sample (*w*< *l*), and $\Delta M$ in unit of emu/cm$^3$ corresponding to the *M-H* loop width between increasing and decreasing field branches. Therefore, the $J_c$ at different temperatures shown in Fig. 8 are obtained. Below 10 K $J_c$ exceeds $10^4$A/cm$^2$, consistent with literature results [34-35].

## IV. CONCLUSIONS

The microstructure evolution and its relationship with superconductivity are investigated in $K_xFe_{1.6+y}Se_2$ single crystals obtained by as-quenching, furnace-cooling, and post-annealing methods, respectively. We demonstrate that the large shielding fraction results from the uniform and contiguous distribution of the superconducting phase in the form of the network pattern, which can be achieved simply by quenching the samples at/above iron vacancy order-disorder transition temperature $T_s$. Furthermore, the resistivity hump and the dip near zero field in MHLs strongly depend on the connection between the superconducting domains. The real superconducting volume fraction is determined by the solubility of excess iron atoms ~$y$ in $K_{0.8}Fe_{1.6}Se_2$ liquid melt. In our results, the starting mixtures $K_{0.8}Fe_{2+z}Se_2$ (0.2<$z$<0.6) yield a real composition close to $K_{0.81}Fe_{1.70}Se_2$.

Upon cooling below $T_s$, the superconducting phase precipitates from $K_xFe_{1.6+y}Se_2$ solid solution, driven by the iron vacancy order–disorder transition. Lattice strains and interface energy should be responsible for the regular arrangement of the superconducting phase along certain crystallographic orientation between the two phases. It is further suggested that charge imbalance in superconducting $K_xFe_2Se_2$ is compensated through lattice strains in phase separated $K_xFe_{1.6+y}Se_2$. By slowly cooling across $T_s$ with an application of uniaxial pressure along the [110] or [1$\bar{1}$0] direction, a large and long superconducting strip is expected.

**ACKNOWLEDGMENTS**

This work was supported by the U.S. Department of Energy, Office of Basic Energy Sciences, Division of Materials Science and Engineering. Ames Laboratory is operated for the U.S. Department of Energy by Iowa State University under Contract No. DE-AC02-07CH11358.

Figure captions

FIG. 1. (Color online) Left: Optical image of $K_xFe_{1.6+y}Se_2$ ingot grown from a nominal composition of $K_{0.8}Fe_{2+z}Se_2$ ($z$=0.6) by furnace cooling, which is 5 cm long and 0.8 cm in diameter. Right top: Shiny surface of the crystals by post-annealing at 873 K for 4 hours. The ruler is in centimeter scale. Right bottom: Plate-like crystal cleaved from above annealed crystal.

FIG. 2. (Color online) (a) XRD patterns show the (00$l$) orientation of $K_xFe_{1.6+y}Se_2$ single crystals grown from the nominal composition of $K_{0.8}Fe_{2+z}Se_2$ ($z$=0.6) by furnace cooling (FC), and post-annealing (PA) at 873 K for 2 and 6 hours, respectively. The shoulder beside the main reflection corresponds to the superconducting phase, as indicated by the asterisk. Temperature dependence of local (008) reflection within the range $59.5°<2\theta<61.5°$ measured upon warming (b) displays that the shoulder merges into the main reflection above 537 K, whereas it comes out below 537 K by cooling (c), measured within the range from 493 to 553 K at an interval of 2 K. For the sake of clarity, XRD patterns were shifted vertically. (d) Temperature dependence of $c$-axis lattice parameters of superconducting phase ($c$1) and antiferromagnetic ordered phase ($c$2).

FIG. 3. (Color online) (a) $K_xFe_{1.6+y}Se_2$ single crystals obtained by as-quenching (AQ) at 1173 K shows $T_c$ at 28.6 K. (b) The crystals obtained by furnace cooling (FC) show very weak superconductivity, whereas the shielding fraction has been greatly enhanced by post-annealing (PA) at 873 K for 2 and 6 hours, and 543 K for 10 minutes. One batch shows $T_c$ at 28.7 K, and the other $T_c$=32 K (c) Normal-state magnetic susceptibility $\chi(T)$ decreases with decreasing temperature for AQ, FC, and PA samples. Samples are from the nominal composition of $K_{0.8}Fe_{2+z}Se_2$ ($z$=0.6).

FIG. 4. (Color online) SEM images reveal the coexistence of two phases in $K_xFe_{1.6+y}Se_2$ single crystals. (a) The crystals obtained by as-quenching (AQ) at 1073 K showing bright network. Inset shows back Laue x-ray diffraction pattern of the same crystal. (b) The simulated back Laue pattern. (c) Schematic drawing of phase separation. Network is supposed to be 122 structure, where iron vacancy sites (Fe2 sites) are filled with red spots. The iron vacancy sites in the $\sqrt{5}\times\sqrt{5}\times1$ superstructure, that is, iron vacancy ordering status (OS), are represented by the cross (Fe2 sites). Fe1 sites (blue spots) are fully occupied. The length of the long side of rectangular bar ranges from several hundred nanometers in as-quenched and post-annealed samples to ~1 μm in furnace-cooled samples. In order to display the crystal structures of the superconducting phase and iron vacancy ordered phase, rectangular bars were drawn at a scale of ~1:100. Lattice distortions in the interface between two phases are omitted. (d) Iron vacancy disordering status (DOS). Network is observed in the crystals obtained by as-quenching at (e) 1173 K and (f) 1073 K. The furnace-cooled (FC) crystals with different starting materials $K_{0.8}Fe_{2+z}Se_2$ (g) $z=0$, (h) $z=0.2$, and (i) and (j) $z=0.6$ under different magnification. Network aggregates into bar chains by slowly cooling across the iron vacancy order-disorder transition temperature $T_s$. Network can be recovered by post-annealing (PA) at 673 K for 2 hours (k), 873 K for 2 hours (l), and 873 K for 6 hours (m)-(n) under different magnification. Network becomes finer in (n) with a modulated wavelength $\lambda<1$ μm. The network disappears and bar chains recover by a secondary heat treatment (2nd HT) in post-annealed samples showing sharp transition and large shielding fraction with nominal compositions $K_{0.8}Fe_{2+z}Se_2$ (o) $z=0.2$ and (p) $z=0.6$.

FIG. 5. (Color online) (a) DSC curve of $K_xFe_{1.6+y}Se_2$ single crystals from the nominal composition of $K_{0.8}Fe_{2+z}Se_2$ ($z=0.6$). Iron vacancy order–disorder transition is clearly observed at $T_s=544$ K upon warming. The heating and cooling cycles were applied to detect any exo- and endothermic effects which are assumed to be related to the formation of exsolution texture. (b) $K_xFe_{1.6+y}Se_2$ single crystals start to melt and decompose above 1175 K. Inset shows the optical image of the tantalum capsule in which the crystal with a mass of 40 mg was sealed. (c)-(e) Schematic drawings of $K_xFe_{1.6+y}Se_2$ single crystals growth from liquid melt. Quenching above $T_s$ gives rise to network with large shielding fraction, whereas slowly cooling yields bar chains with small shielding fraction, which strongly suggests solid solution exists above $T_s$ and phase separation occurs below $T_s$.

FIG. 6. (Color online) (a) Temperature dependence of in-plane resistivity $\rho_{ab}$ of $K_xFe_{1.6+y}Se_2$ single crystals obtained by as-quenching (AQ), furnace cooling (FC), and post-annealing (PA). The resistive broadening curves of $K_xFe_{1.6+y}Se_2$ single crystals under different magnetic fields for both (b) $H//c$ and (c) $H//ab$ configurations.

FIG. 7. (Color online) Magnetic hysteresis loops (MHLs) of $K_xFe_{1.6+y}Se_2$ single crystals obtained by post-annealing at 873 K for 4 hours for (a) $H//c$ and (b) $H//ab$ at different temperatures.

FIG. 8. (Color online) The critical current density $J_c$ estimated from magnetization curves by using Bean's extended critical state model.

TABLE I. Preparation and heat treatment conditions.

| Crystal | Preparation conditions | | | | | Post annealing[3] | | Secondary heat treatment[4] |
|---|---|---|---|---|---|---|---|---|
| | Starting mixture | Soaking temperature | Soaking time | Growth cooling rate | Post growth cooling | Annealing temperature | Annealing time | |
| S1 | K0.8Fe2.0Se 2.0 | 1303 K | 2h | 6 K/h to 1053 K | furnace cooling[1] from 1053 K | 673 K | 2, 4, 6h | |
| | | | | | | 873 K | 2, 4, 6h | |
| S2 | K0.8Fe2.2Se 2.0 | 1303 K | 2h | 6 K/h to 1053 K | furnace cooling from 1053 K | 673 K | 2, 4, 6h | a |
| | | | | | | 873 K | 2, 4, 6h | |
| | | | | | | 543 K | 10 min | |
| S4 | K0.8Fe2.6Se 2.0 | 1303 K | 2h | 6 K/h to 1053 K | furnace cooling from 1053 K | 673 K | 2, 4, 6h | b |
| | | | | | | 873 K | 2, 4, 6h | |
| | | | | | | 543 K | 10 min | |
| S19 | K0.8Fe2.6Se 2.0 | 1403 K | 2h | 5 K/h to 1173 K | quenching[2] from 1173 K | | | |
| S20 | K0.8Fe2.6Se 2.0 | 1403 K | 2h | 5 K/h to 1073 K | quenching from 1073 K | | | |

1) Furnace was naturally cooled down to 373 K during a period of 12 hours by switching off the power.

2) Samples were quenched in room temperature water.

3) Samples were quenched in room temperature water after annealing.

4) (a) Samples annealed at 873 K for 2 hours and (b) samples annealed at 873 K for 2 hours were heated to 873 K and naturally cooled down to room temperature by switching off the power.

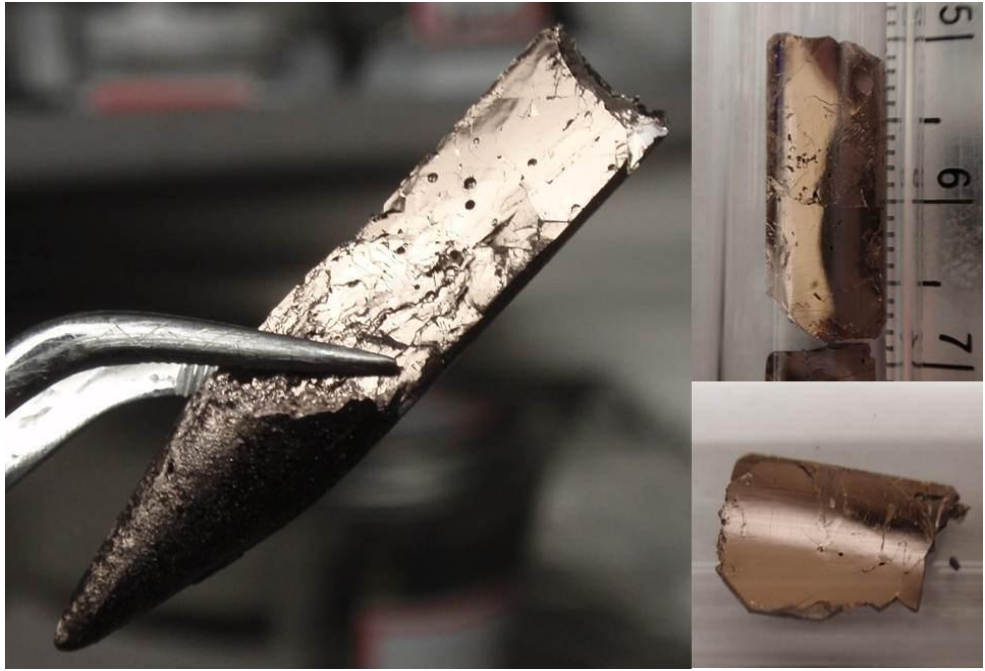

Figure 1

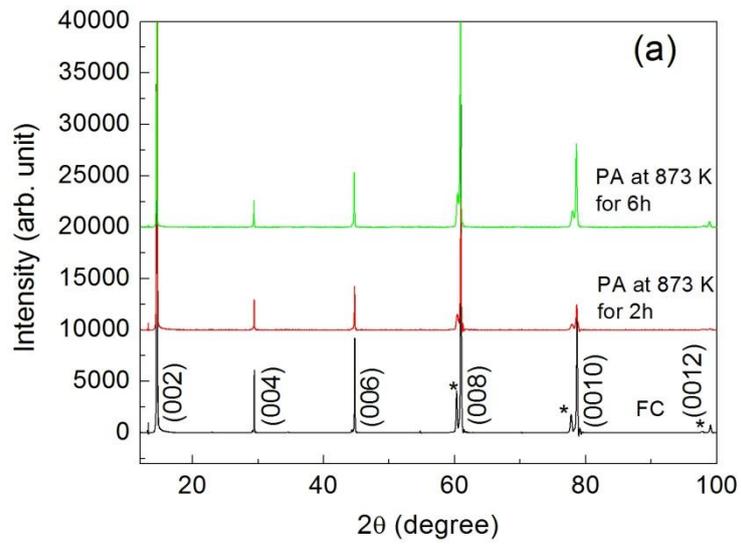

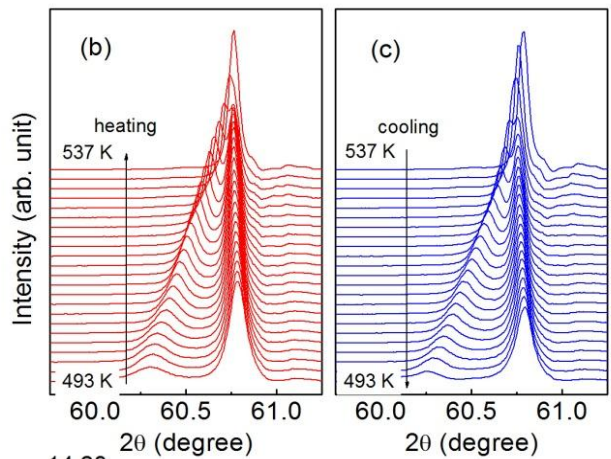

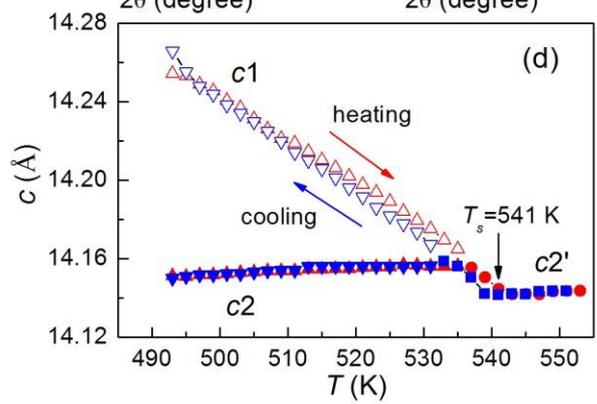

Figure 2

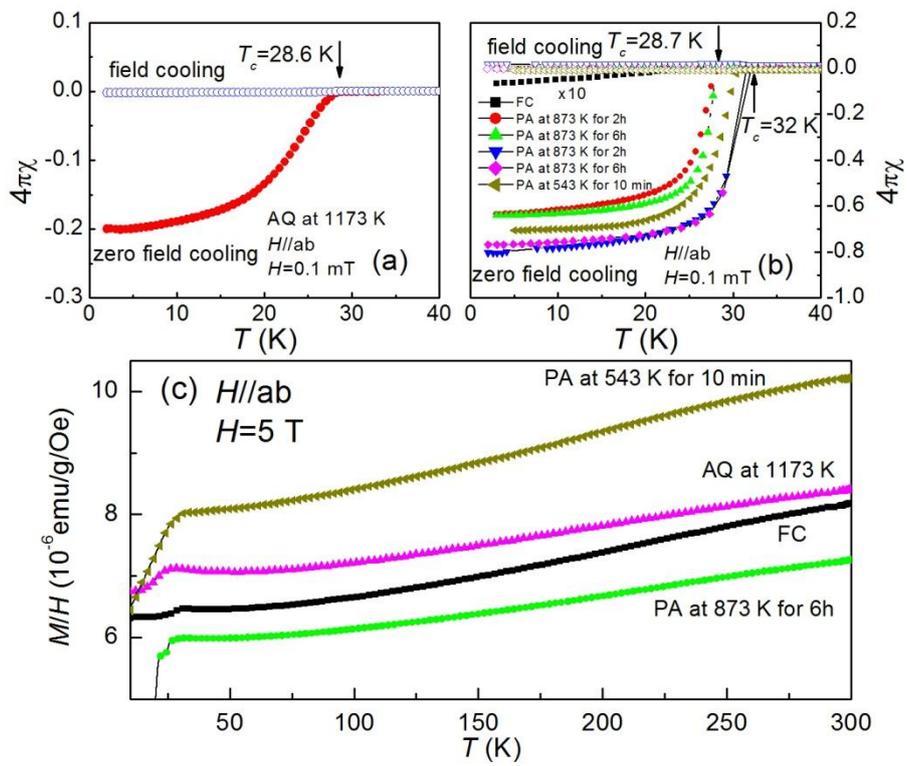

Figure 3

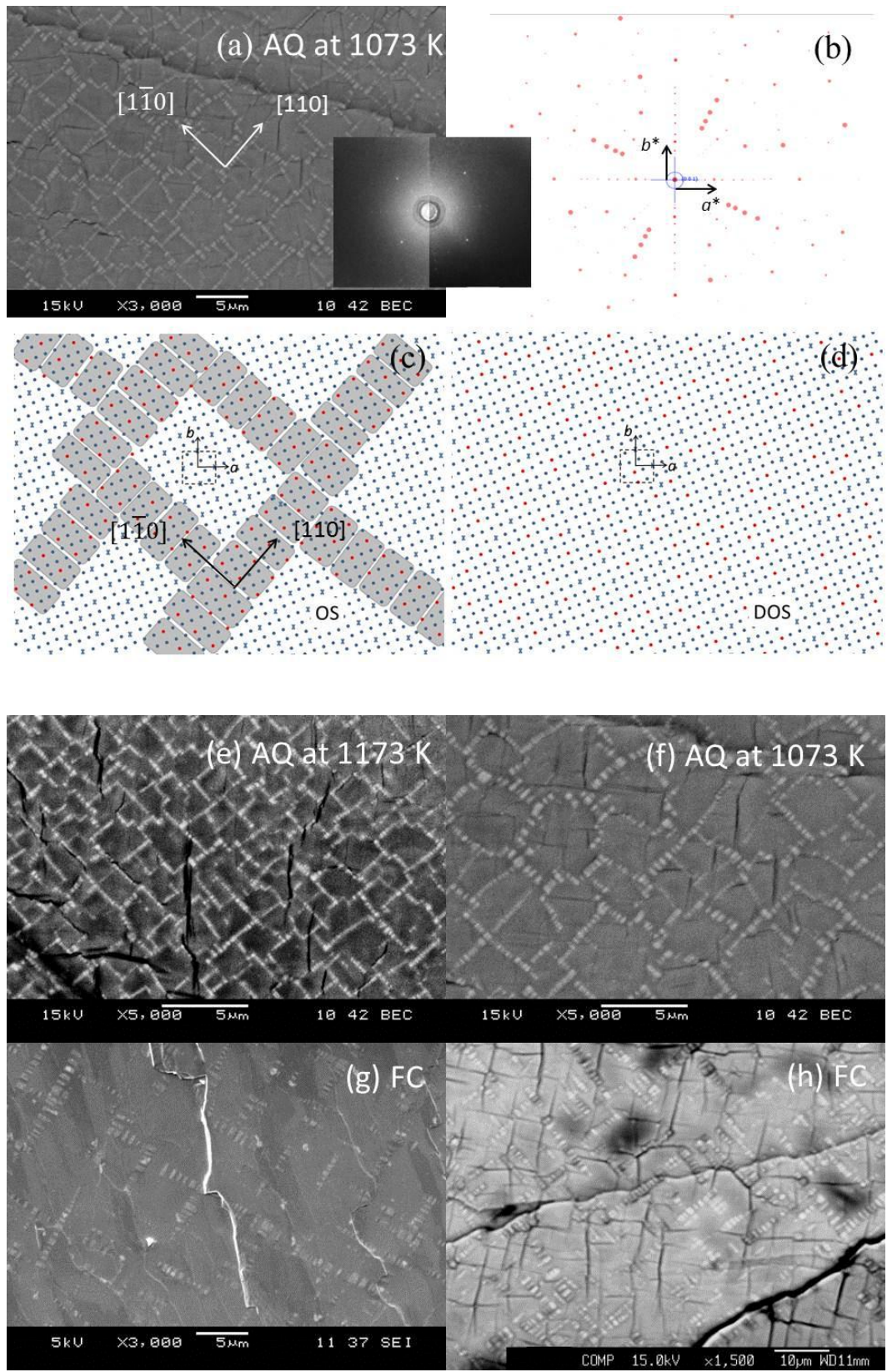

Figure 4

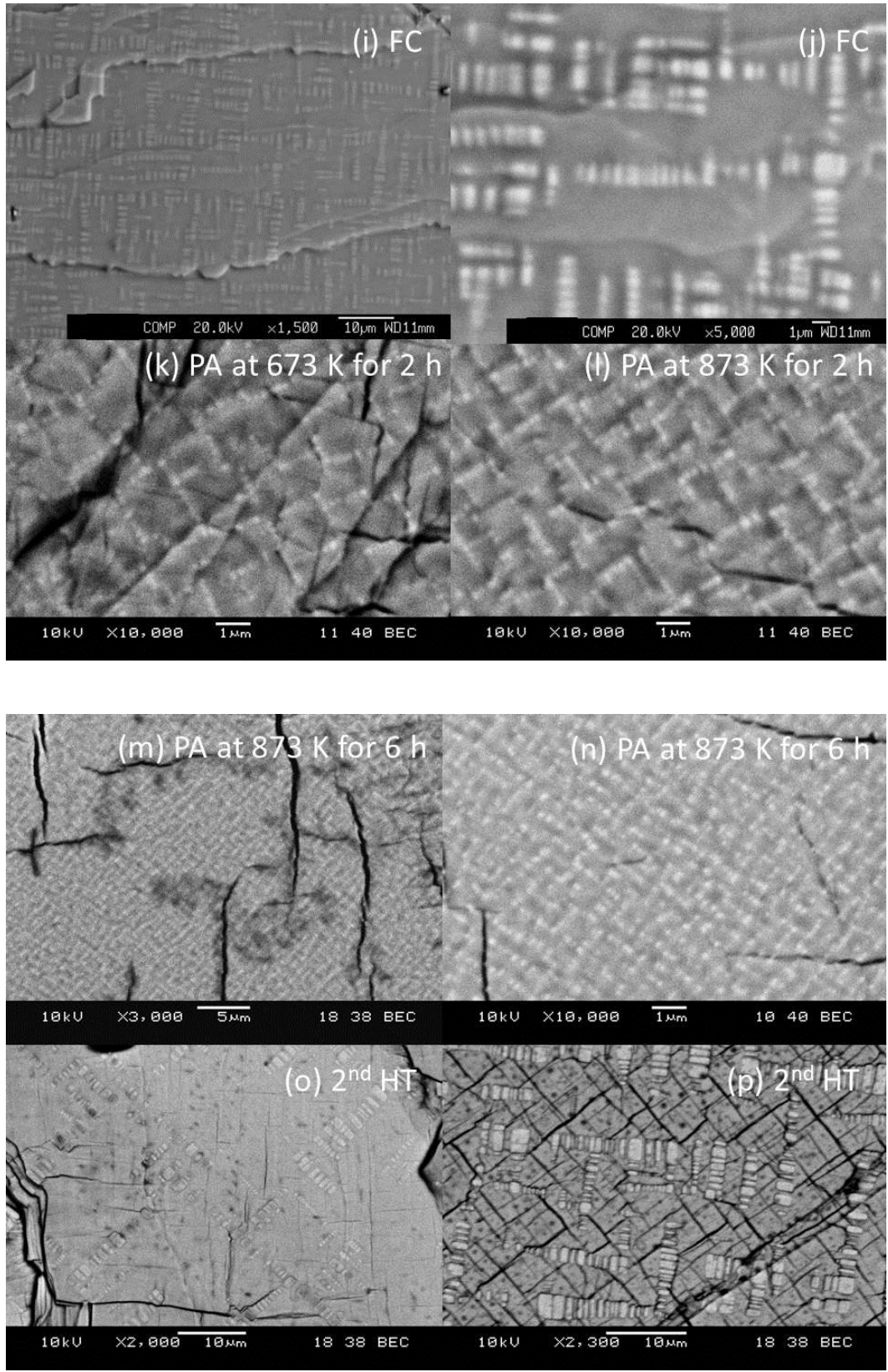

Figure 4

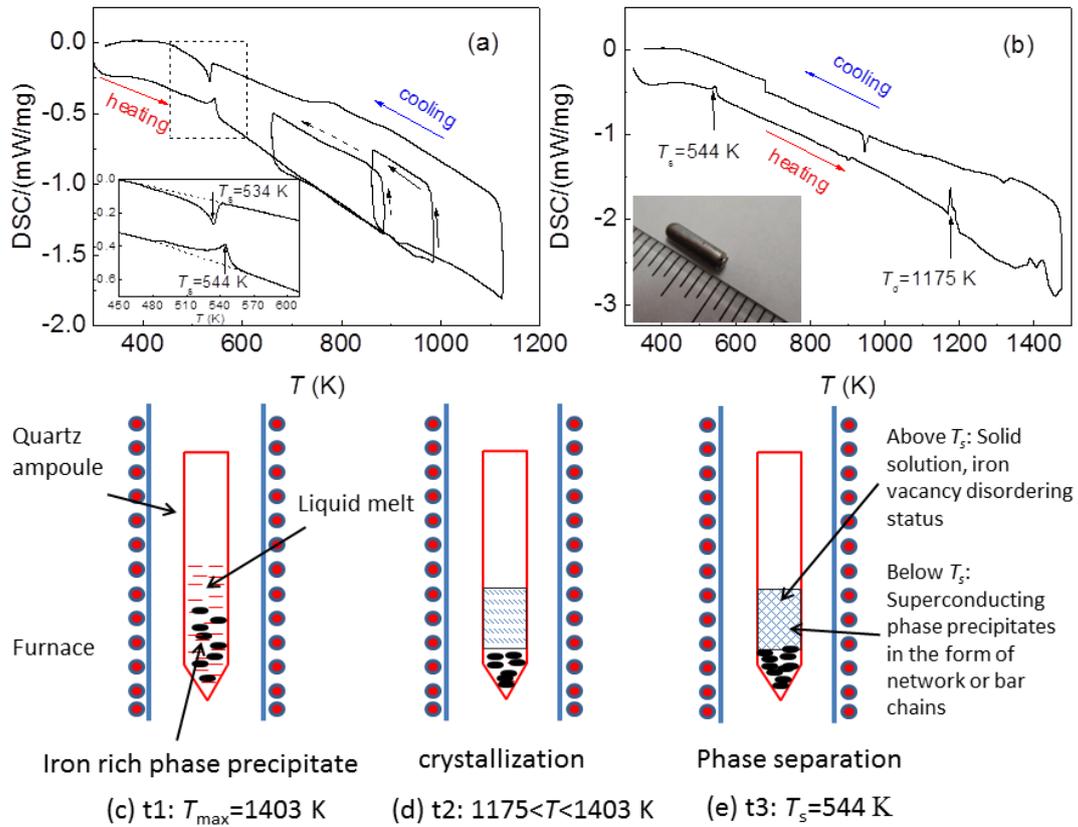

Figure 5
27

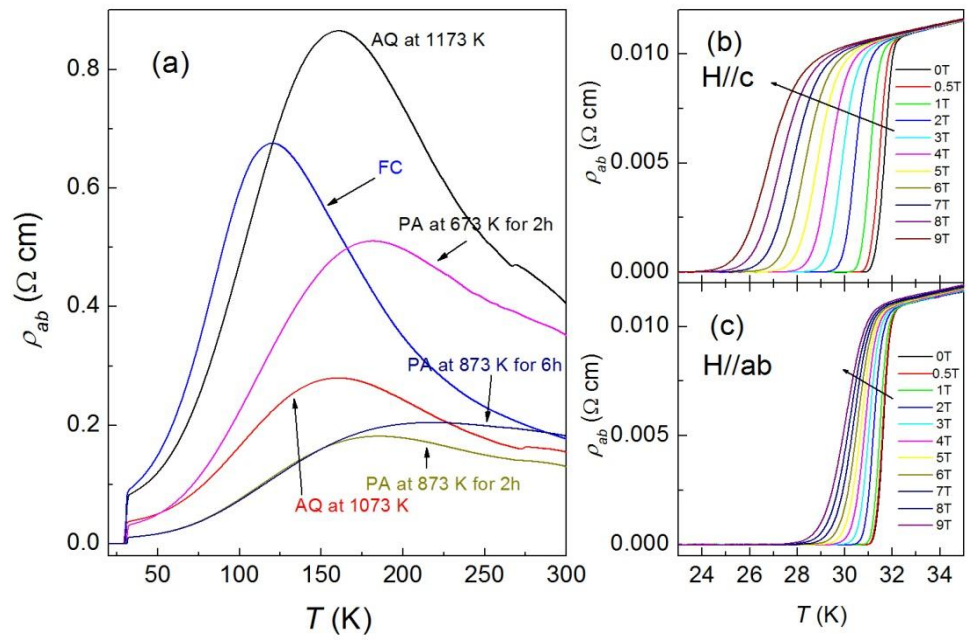

Figure 6

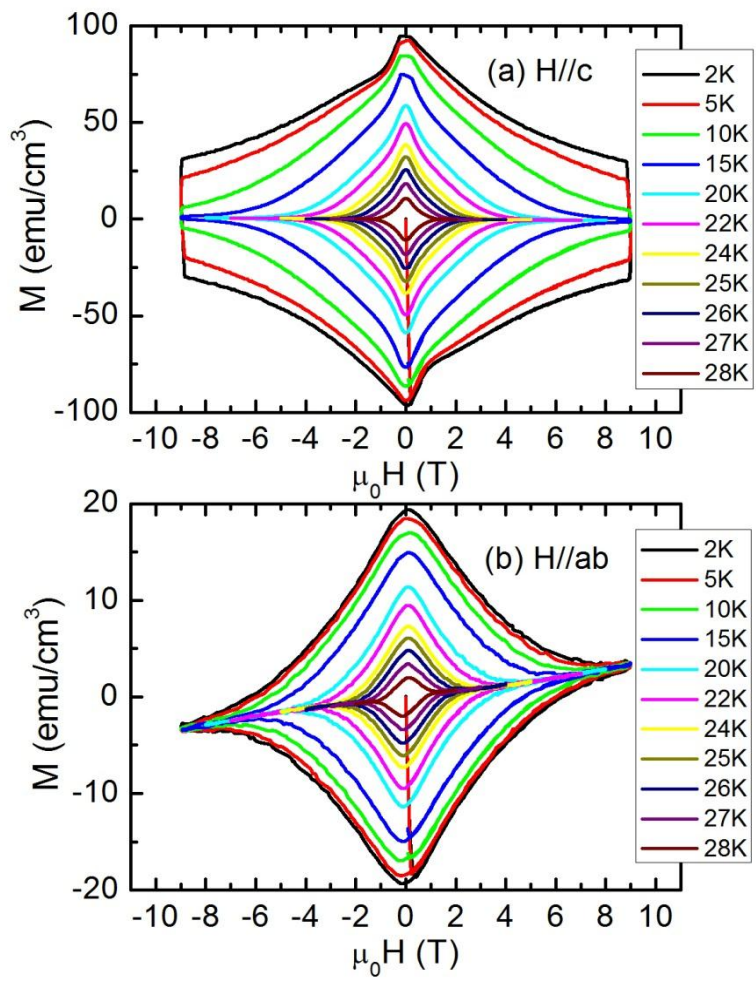

Figure 7

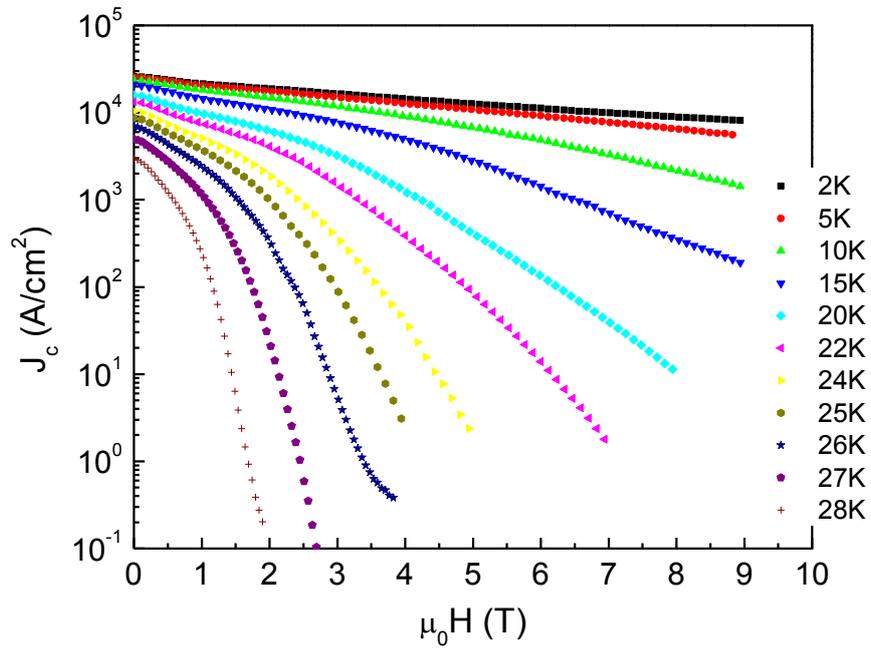

Figure 8